\documentclass[aps,prc,twocolumn,showpacs]{revtex4}
\usepackage{ulem}
\usepackage{color}
\usepackage{graphicx}
\usepackage{epsfig}
\usepackage{bm}
\def\be {\begin{equation}}
\def\ee {\end{equation}}
\def\nn {\nonumber}
\def\bea {\begin{eqnarray}}
\def\eea {\end{eqnarray}}

\def\del{\partial}
\newcommand{\ep}{\epsilon}
\newcommand{\om}{\omega}  
\newcommand{\vk}{\vec k}
\newcommand{\vq}{\vec q}

\begin{document}


\title{Finite size effect on Dissociation and Diffusion of chiral partners in Nambu-Jona-Lasinio model}
\author{Paramita Deb$^1$, Sabyasachi Ghosh$^2$, Jai Prakash$^3$, Santosh Kumar Das$^3$, Raghava Varma$^1$}

\affiliation{$^1$ Department of Physics, Indian Institute of Technology Bombay, Powai, Mumbai- 400076, India}
%
\affiliation{$^2$ Indian Institute of Technology Bhilai, GEC Campus, Sejbahar, Raipur-492015, 
Chhattisgarh, India}
%
\affiliation{$^3$ School of Physical Science, Indian Institute of Technology
Goa, Ponda-403401, Goa, India}
%

\begin{abstract}
Along with masses of pion and sigma meson modes, their dissociation into quark medium provide a detail
spectral structures of the chiral partners. Present article has studied a finite size effect on that
detail structure of chiral partners by using the framework of Nambu-Jona-Lasinio model. Through this
dissociation mechanism, their diffusions and conductions are also studied. The masses, widths,
diffusion coefficients, conductivities of chiral partners are merged at different temperatures in restore phase
of chiral symmetry, but merging points of all are shifted in lower temperature, when one introduce finite
size effect into the picture. The strengths of diffusions and conductions are also reduced due to finite size
consideration.

\end{abstract}

\maketitle
\section{Introduction}

The nuclear matter, formed in the heavy-ion collision experiments like relativistic heavy ion collider (RHIC)
and large hadron collider (LHC), has a finite volume and life time in fm and fm/c scale respectively.
Initially a hot quark gluon plasma (QGP) is expected to be form, then it will expand and after a particular
volume and time, the medium will freeze out. 
Experimentally, this freeze out volume can be measured via 
Hanbury-Brown-Twiss (HBT) methodology, whose details can be found in review articles~\cite{HBT_rev1,HBT_rev2} 
and references therein. This freeze out volume of the matter depending on the size 
of the colliding nuclei, center of mass energy and collision centrality~\cite{RHIC_BES}. 
A freeze out volume range 2000-3000 fm$^3$ within a large range of center of mass energy $\sqrt{s}$
is shown by Ref.~\cite{HBT_pi}. On the other hand, volume range 50-250 fm$^3$ is expected from
Refs.~\cite{UrQMD1,UrQMD2}. So it is important to understand different Phenomenological quantities if they have any finite size effect 
within this uncertain volume range of RHIC or LHC matter.

The effects of finite volume have been addressed by many models such as 
non-interacting bag model \cite{elze},
quark-meson (QM) model \cite{Braun1,Braun2,Braun3,Braun4,Braun5,Fraga1,Braun6}, 
Nambu--Jona-Lasinio (NJL) model \cite{Hosaka,Abreu1,Elbert1,Abreu2,Abreu3,Abreu4,NJLV_2018,Abreu_uds},
Polyakov loop extended NJL (PNJL)~\cite{Deb1,Sur,Kinkar1,SG1}, 
Polyakov loop extended linear sigma model (PLSM)~\cite{Magdy_17}, 
hadron resonance gas (HRG)~\cite{Greiner,Samanta1,Xu_PLB,SG2,SG3,SG_HRGV_rev,Deb_HRG},
Walecka model~\cite{Abreu5,Casimir} etc. All of the model calculations accept that the 
quark-hadron phase diagram can depend on the volume of the matter. Among the NJL and PNJL model
studies~\cite{Hosaka,Abreu1,Elbert1,Abreu2,Abreu3,Abreu4,NJLV_2018,Abreu_uds,Deb1,Sur,Kinkar1,SG1} 
on finite size effect, only Ref.~\cite{NJLV_2018} have recently studied on finite size effect of meson 
masses, whose more detail extension might provide a rigorous understanding on the properties of chiral
partners - $\pi$ and $\sigma$ mesons. Along with the masses of $\pi$ and $\sigma$ mesons, their dissociation
probability into quark medium will give a full spectral details of them. Finite size effect on this detail
spectral structure of the chiral partners is not been studied before and present work has attempted this task.

The article is organized as follows. Next in Sec.~(\ref{sec:Form}), we have built Formalism part, which carry
3 subsections. First in Sec.~(\ref{sec:NJL}), framework of Nambu-Jona-Lasinio model with finite size is
introduce, then in Sec.~(\ref{sec:mass}), the expression of mesonic masses and decay widths are derived,
and at the end of Formalism part, in Sec.~(\ref{sec:diff}), framework of diffusion and conductivity are
constructed for mesonic modes. After addressing formalism part, we have explored the numerical results of
masses, decay widths, diffusion coefficients, conductivities of the mesonic modes and their finite size effect.
Finally, we have summarized our investigation in Sec.~(\ref{sec:sum}).

\section{Formalism}
\label{sec:Form}
\subsection{NJL model at finite volume}
\label{sec:NJL}
\begin{figure} 
\includegraphics[scale=0.5]{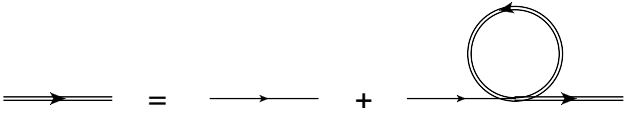}
\caption{Diagrammatic representation of Gap Eq.~(\ref{NJL1}), which make connection between
dressed quark (double solid line) with constituent mass $M_f$ and bare quark (single solid line) 
with current quark mass $m_f$} 
\label{Gap}
\end{figure}

We consider the framework of the Nambu--Jona-Lasinio (NJL) model \cite{hatsuda,hansen} 
for the description of the coupling between quarks and the chiral condensate 
in the scalar-pseudo-scalar sector. This model nicely captures the chiral 
symmetry and its spontaneous breaking physics. We will use a two-flavor model, 
with a degenerate mass matrix for $u$ and $d$ quarks. The Lagrangian can 
be written as
\begin{eqnarray}
   {\cal L} &=& {\sum_{f=u,d}}{\bar\Psi_f}\gamma_\mu i\partial^\mu
             {\Psi_f}-\sum_f m_{f}{\bar\Psi_f}{\Psi_f}\nonumber\\
       &+& G[\sum_f ({\bar\Psi_f} {\Psi_f})^2+
            ({\bar\Psi_f} i\gamma_5\tau^a {\Psi_f})^2] 
\end{eqnarray}  
where $f$ denotes the flavors $u$, $d$ respectively, $m_f=m_u=m_d$, $\tau^a$
are $SU_f(2)$ Pauli matrices acting in flavor space. 
%
%
As a result of dynamical breaking of chiral
symmetry in the NJL model, the chiral condensate $\langle\bar{\Psi}\Psi\rangle$
acquires non-zero vacuum expectation values. The constituent mass
as a consequence is given by,
\begin{equation}
 M_f~=~m_f-2G\sigma_f
 \label{NJL1}
\end{equation}
where $\sigma_f\equiv\langle \bar{\Psi}_f\Psi_f \rangle$ represents the 
chiral condensate. Its diagrammatic representation is sketched in Fig.~(\ref{Gap}). 

Now in order to implement the effect of finite system sizes, one is ideally supposed 
to choose the proper boundary conditions : periodic for bosons and anti-periodic for
fermions. This in effect leads to a sum of infinite extent over discretized 
momentum values, $p_i=\frac{\pi n_i}{R}$, $R$ being the dimension of cubical volume.
$n_i$ are positive integers with i=x,y,z. This would then imply as lower momentum cut-off
$p_{min}=\frac{\pi}{R}=\lambda (\rm say)$. 
%
The infinite sum over discrete momentum values
will be replaced by integration over continuum momentum variation, albeit with the 
lower momentum cut-off. 
This in effect implies that the system volume, $V$ will be
regarded as a parameter just like temperature, T and chemical potential, $\mu$ on the 
same footing. Parametrization will be the same as for zero T, zero $\mu$ and infinite V.
Any variation therefore occurring due to any of these parameters will be reflected in
$\sigma_f$, $\Phi$ etc. and through them in meson spectra.

With these simplifications, the thermodynamic potential thereafter takes the form,
\begin{eqnarray}
\Omega &=& -{\frac {2 N_c N_f}{(2 \pi)^3}} \int {\sqrt {p^2 + M^2}} dp\nonumber\\
         &-&{\frac{2 N_c N_f T}{(2 \pi)^3}} \int dp (ln(1+exp(-\frac{(E-\mu)}{T}))\nonumber\\
         &+& ln (1+ exp (-\frac{(E+\mu)}{T}))) + \frac{(M_f- m_f)}{4G}
\end{eqnarray}
where, each term bears its usual significance, which can be found 
in~\cite{Saumen}. The parameters used in eq. (1) are usually fixed to 
reproduce the mass and 
decay constant of the pion as well as the chiral condensate. The parameters
are given in Table~(\ref{table1}).

\begin{table} 
\begin{center}
\begin{tabular}{|c|c|c|c|c|c|c|c|c|c|c|}
\hline
$ m_u  $ & $ \Lambda  $ & $ g_S \Lambda^2 
$ & $|\langle \bar\Psi_u \Psi_u\rangle|^{1\over 3}  $ &$ f_\pi
$ & $ m_\pi $ \\ 
$(MeV) $ & $ (GeV) $ & $ (GeV^-2) $ & $ (MeV) $&$ (MeV) $ & $ (MeV) $\\
\hline
$ 5.5 $&$ 0.651 $&$ 5.04  $&$ 251 $&$ 92.3 $&$139.3 $ \\
\hline
\end{tabular}
\caption{Parameters of the Fermionic part of the model.}  
\label{table1}
\end{center}
\end{table}  
\subsection{Mesonic Excitations}
\label{sec:mass}
\begin{figure} 
\includegraphics[scale=0.5]{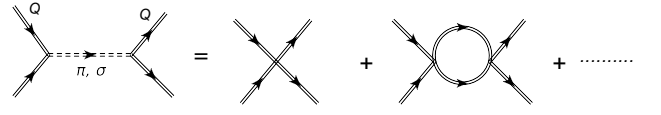}
\caption{Diagrammatic representation of transformation of quark-quark interaction into a effective
$\pi$, $\sigma$ meson propagators (double dash line), given in Eq.~(\ref{D_M}).} 
\label{QM}
\end{figure}
The properties of the medium beyond bulk thermodynamic properties 
can be understood by studying the low-lying mesonic excitations. The 
masses and decay widths of the mesonic resonances are calculated from the 
correlations of ${\bar \psi}_f \Gamma \psi_f$ type operators in QCD vacuum. The masses
and the spectral functions of pseudo-scalar and scalar mesonic states are 
interesting because of their close connection with the chiral symmetry 
breaking and its restoration. The collective excitations, that is, the 
fluctuation of the mean field around the vacuum, can be handled within 
the random-phase-approximation (RPA)~\cite{Saumen}. Its schematic Feynman diagram~\cite{Buballa}
is shown in Fig.~(\ref{QM}), revealing the transformation of quark-quark interaction into a effective
mesonic propagators. 
In this approximation, the
retarded correlation function is given by
\begin{equation}
 D^R_M (\omega,\overrightarrow{p}) = \frac {2iG} {1-2 G \Pi_M (\omega, \overrightarrow{p})} 
 \label{D_M}
\end{equation}
$\Pi_M (p^2)$ is the one loop polarization function for the mesonic channel
under consideration, 
\begin{equation}
 \Pi_M (p^2) = \int {\frac {d^4p} {(2\pi)^4}} Tr [\Gamma_M S({\frac {(q+p)} {2}}) \Gamma_M ({\frac {(q- p)} {2}})]
\end{equation}
$S(q)$ is the quark propagator and $\Gamma_M$ is the effective vertex factor. The mass of the 
meson is extracted from the pole of the meson propagator at zero momentum, given by the equation 
\begin{equation}
 1- 2G Re \Pi_M (m_M, \overrightarrow{0}) = 0
\label{polemass}
\end{equation}
The mass of the unbound resonance has been considered as the real part of $\Pi_M$. For 
bound state solutions $(\omega =m_M < 2M)$, the polarization function is always real. 
For $m_M > 2M$, $\Pi_M$ has an imaginary part and the meson spectral function gets
a continuum contribution. The meson is no longer a bound state but a resonant one. 
If $Im\Pi$ stays constant around the position of the peak, the spectral function will
be approximated by a Lorentzian with a decay width
\begin{equation}
\gamma_M = 2G Im \Pi_M (m_M, \overrightarrow{0}).
\label{decay-width}
\end{equation}
Explicitly,
\begin{eqnarray}
 \Pi_\pi (m_\pi,\overrightarrow{0}) = I_1 - m^2_\pi I_2 (m_\pi,0)\nonumber\\
 \Pi_\sigma (m_\sigma, \overrightarrow{0}) = I_1 - (m^2_\sigma - 2 M^2) I_2 (m_\sigma)
\end{eqnarray}
where
\begin{equation}
 I_1 = \frac {2 N_c N_f} {(2\pi)^3} \int \frac{dq}{E_q} (1- f_-(\overrightarrow{q},\beta,\mu)- 
     f_+(\overrightarrow{q},\beta,\mu))
\end{equation}
and 
\begin{eqnarray}
 I_2(m_{\pi / \sigma}) &=&\frac{2 N_c N_f}{(2 \pi)^3} \int \frac{dq}{E_q}
 (1- f_{-}(q, \beta, \mu) - f_{+}(q,\beta, \mu))
 \nonumber\\
 &&~~~~~~~~ \frac{1}{m^2_{\pi / \sigma} - 4E^2}~.
\end{eqnarray}
The masses of the pion and sigma mesons are given, using the gap equation
\begin{eqnarray}
 {m_0 \over M + 2 G m^2_\pi Re I_2 (m_\pi)} = 0
\end{eqnarray}
and
\begin{equation}
 {m_0 \over M + 2 G (m_\sigma^2 - 4 M^2) Re I _2 ( m_\sigma)} =0
\end{equation}
The real and imaginary part of $\Pi_M (\omega, \vec{0})$ are given as 
\bea
 Re\Pi_M (\omega, \vec{0})&=& {\frac{2 N_c N_f}{(2 \pi)^3}} \int d\vec {q} {\frac{1}{E_{\vec q}}}
 {\frac{{E^2_{\vec q}} - {\epsilon_M}/4}{{E^2_{\vec q}}- {\omega^2/4}}}
 \nn\\
 &&  (1-f_{-}(E_{\vec q})-f_{+}(E_{\vec  q}))
 \label{Re_M}
 \eea
 \bea
 Im\Pi_M(\omega, \vec {0}) &=& \theta (\omega^2 - 4m^2) {\frac{N_c N_f}{8\pi \omega}}(\omega^2 - \epsilon_M^2)\nonumber\\
 &&(1-f_{-}(\omega)-f_{+}(\omega))~.
 \label{IM_M}
\eea
Here, $f_{\mp}(x)$ is the Fermi distribution function for particles and antiparticles. 

\subsection{Diffusion of chiral partners}
\label{sec:diff}
We know that $\pi$ and $\sigma$ meson are pseudo-scalar and scalar
modes of same quark condensate with same spin quantum no ($J=0$)
but different parity states $\pi=-1, +1$. Particle physicist generally
follow the compact notation of Spin and parity quantum
no as $J^\pi$. So we can visualize that $\pi$ and $\sigma$ meson are
basically $J^\pi=0^{-1}$ and $0^{+1}$ quantum states of same quark composition.
They are called as chiral partners. Quarks are the fundamental
building block of hadrons and a non-zero quark condensate is responsible for 
the mass difference between chiral partners at $T=0$. Now, increasing the 
temperature of nuclear or hadronic matter, a phase transition from hadrons to quarks
can be occurred beyond a transition temperature and corresponding condensate
also melts down. Due to this transformations - from non-zero to zero condensate,
from breaking to restore phase of chiral symmetry, the mass difference of chiral
partners are disappeared. NJL model (as well as other effective QCD models) can 
nicely capture these facts, whose mathematical framework is already addressed
in earlier Secs.~(\ref{sec:NJL}) and (\ref{sec:mass}) and results will be discussed
in next Sec.~(\ref{sec:Res}). In present subsection, we will discuss about the 
drag, diffusion and conduction of two mesonic modes via dissociations into quarks and
anti-quarks. The mathematical anatomy of dissociation process, already addressed in
Eq.~(\ref{IM_M}), which can directly be connected with drag and then with 
diffusion and conduction of chiral partners. Here we will build first the 
diagrammatic construction of conductivity and diffusion. Then, at the end, we
will see the position of drag in the diagrammatic expression of conductivity
and diffusion.
\begin{figure} 
\includegraphics[scale=0.5]{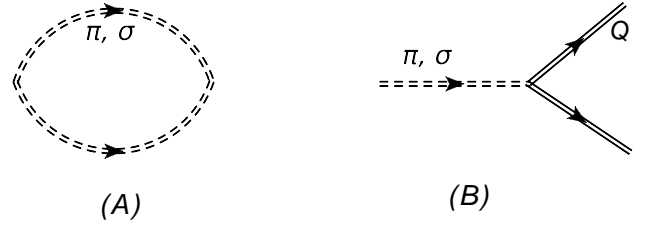}
\caption{(A) One-loop schematic representation of $\pi$ or $\sigma$ meson current-current correlator, whose
low frequency limit is connected with their diffusion coefficient or conductivity. (B) Dissociation
diagram of $\pi$ or $\sigma$ meson to quark and anti-quark.} 
\label{Diff}
\end{figure}

In real-time thermal field theory, two point
function of meson current ($J_\mu\equiv \phi\del_\mu \phi$) 
can be expressed in a $2\times 2$ matrix structure,
which can normally be diagonalized in terms of a single element 
like retarded component $\Pi^R_{\mu\nu}(q_0,\vq)$ or the spectral
function $\rho_{\mu\nu}(q_0,\vq)$ of that mesonic correlator.
Starting with 11 component ($\Pi^{11}_{\mu\nu}$) of the $2\times 2$ matrix, 
one can obtain anyone of these quantities
by using of their connecting relation:
\bea
\rho_{\mu\nu}(q_0,\vq)&=&2{\rm Im}\Pi^R_{\mu\nu}(q_0,\vq)
\nn\\
&=&2{\rm tanh}(\frac{\beta q_0}{2}){\rm Im}\Pi^{11}_{\mu\nu}(q_0,\vq)~.
\label{R_bar_11}
\eea
With the help of Wick contraction technique,
the 11 component of current-current correlator 
can be derived as
\bea
\Pi^{11}_{\mu\nu}(q_0,\vq)&=&i\int d^4x e^{iqx}\langle T J_\mu(x)J_\nu(0) \rangle_\beta
\nn\\
&=&i\int d^4x e^{iqx}
\langle T{\phi}\underbrace{(x)\del_\mu{\phi}\overbrace{(x)
{\phi}}(0)\del_\nu\phi}(0)\rangle_\beta
\nn\\
&=&i\int \frac{d^4k}{(2\pi)^4}N_{\mu\nu}(q,k)
D_{11}(k)D_{11}(p=q+k)~,
\nn\\
\label{P_11}
\eea
where
\be
D^{11}(k)=\frac{-1}{k_0^2-\om_k^2+i\ep}+2\pi i n_k \delta(k_0^2-\om_k^2)~,
\label{E11}
\ee
where $n_k=1/(e^{\beta\om_k} - 1)$
is Bose-Einstein (BE) distribution function of meson
and
\be
N_{\mu\nu}(q,k)=-4k_\mu(q-k)_\nu~.
\label{N_qk}
\ee
The Eq.~(\ref{P_11}) can diagrammatically be associated with a one-loop kind of 
self-energy diagram with meson internal lines, shown in Fig.~\ref{Diff}(A).
Now from this correlator $\Pi^{11}_{\mu\nu}$ one can identify the useful correlators - 
(1). density-density correlator: 
\be
\Pi^{R}_{00}(q_0,\vq)=i\int d^4x e^{iqx}\langle [J_0(x),J_0(0)]\rangle_\beta
\ee
and (2). spatial current-current correlator:
\be
\Pi^{R}_{ij}(q_0,\vq)=i\int d^4x e^{iqx}\langle [J_{ij}(x),J_{ij}(0)]\rangle_\beta~.
\ee
The spatial current-current correlator can be decomposed in transverse and longitudinal
components as
\be
\Pi^{R}_{ij}(q_0,\vq)=\left(\frac{q_iq_j}{q^2} -\delta_{ij} \right)\Pi^{R}_{T}(q_0,\vq)
+ \frac{q_iq_j}{q^2}\Pi^{R}_{L}(q_0,\vq)~,
\ee
where our matter of interest is on longitudinal component $\Pi^{R}_{L}(q_0,\vq)$, which can
be extracted from both density-density and (spatial) current-current correlators by using
the relation:
\be
\Pi^{R}_{L}(q_0,\vq)=\frac{q_0^2}{q^2}\Pi^{R}_{00}(q_0,\vq)=\frac{q^iq^j}{q^2}\Pi^{R}_{ij}(q_0,\vq)~.
\ee
So we can define a spectral function without Lorentz indices
\be
\rho=2\Pi^{R}_{L}(q_0,\vq)~.
\label{rho}
\ee
Using (\ref{E11}) in Eq.~(\ref{P_11}) and then using the other Eqs.~(\ref{R_bar_11}),
(\ref{rho}), we get the simplified structure in positive but low $q_0$ region
\bea
&&\rho(q_0,\vq)=2\int\frac{d^3k}{(2\pi)^3}
\frac{(-\pi)N}{4\om_k\om_p}\{C_3\delta(q_0+\om_k-\om_p)\}
\nn\\
&&~~~~~=2\int\frac{d^3k}{(2\pi)^3}
\frac{N}{4\om_k\om_p}\lim_{\gamma \rightarrow 0}\left[
\frac{C_3\gamma}{(q_0+\om_k-\om_p)^2+\gamma^2}\right]~,
\nn\\
\label{el_G}
\eea
where $C_3\approx q_0\beta\{n^+_k(1-n^+_k)\}$ in the region of $q_0<<\om_k$.

We will take finite value of $\gamma$ in our further
calculations to get a non-divergent values of pion ($\pi$) and sigma ($\sigma$) 
meson conductivity
\bea
\sigma&=&\frac{1}{6}\lim_{q_0,\vq \rightarrow 0}\frac{\rho(q_0,\vq)}{q_0}
\nn\\
&=&\frac{1}{\gamma}2\beta\int^{\infty}_{0} 
\frac{d^3\vk}{(2\pi)^3}\frac{1}{3}\Big(\frac{\vk}{\om_k}\Big)^2[n_k\{1+n_k\}]
\label{cond_ex}
\eea
where 
\be
\chi_s=2\beta\int^{\infty}_{0} 
\frac{d^3\vk}{(2\pi)^3}[n_k\{1+n_k\}]
\ee
is static susceptibility and $\om_k=\{\vk^2+m_{\pi,\sigma}^2\}^{1/2}$. 
So identifying $\gamma_{\pi,\sigma}$ as drag coefficient 
of $\pi$ and $\sigma$ mesons, one can calculate spatial diffusion constant
\be
D=\sigma/\chi_s
\label{D_ex}
\ee

Eq.~(\ref{el_G}) can be approximated as
\be
\sigma=\frac{1}{\gamma}\langle v^2/3\rangle\chi_s~,
\ee
and then
using the further approximated relation $\langle v^2/3\rangle=T/m$, based on 
non-relativistic equipartition theorem, we can easily find Einstein relation 
\be
\langle v^2/3\rangle\frac{1}{\gamma}=\frac{T}{m_{\pi,\sigma}\gamma}=D~.
\label{Einstein}
\ee
The Eqs.~(\ref{cond_ex}) to (\ref{Einstein}) represent basically general
connection among the quantities - $D$, $\gamma$, $\sigma$, $\chi_s$.
Here our interest on those quantities for $\pi$ and $\sigma$ mesonic condensates,
so we can denote them as $D_{\pi,\sigma}$, $\gamma_{\pi,\sigma}$, $\sigma_{\pi,\sigma}$, 
$\chi^{\pi,\sigma}_s$ respectively. Reader can find similar kind of calculation 
of same quantities for heavy quark in Ref.~\cite{Teany}.

Here, we want to obtain drag ($\gamma_{\pi,\sigma}$), diffusion ($D_{\pi,\sigma}$) 
coefficients and conductivity ($\sigma_{\pi,\sigma}$) for $\pi$ and $\sigma$ mesons 
near and beyond Mott temperature,
where those mesonic condensates face mostly quarks and anti-quarks in the medium.
So, it is via $\pi/\sigma\rightarrow Q + {\bar Q}$ decay process, the mesonic condensates
will dissipate through medium, whose corresponding $\gamma_{\pi,\sigma}$, $D_{\pi,\sigma}$
and $\sigma_{\pi,\sigma}$ are our matter of interest to estimate.
Here, drag coefficients of $\pi,\sigma$ states is estimated through the 
decay process of $\pi\rightarrow Q{\bar Q}$ and $\sigma\rightarrow Q{\bar Q}$.
Hence we can exactly equate $\gamma_{\pi,\sigma}$ with Eq.~(\ref{decay-width}),
which estimate decay probability of $\pi, \sigma\rightarrow Q{\bar Q}$ from imaginary
part of mesonic self-energy. This dissociation diagrams are sketched in Fig.~\ref{Diff}(B).
After obtaining the drag coefficients $\gamma_{\pi,\sigma}$, 
we can calculate $D_{\pi,\sigma}$ and $\sigma_{\pi,\sigma}$
with the help of Eqs.(\ref{cond_ex}), (\ref{D_ex}).

%
%

\section{Results and discussion}
\label{sec:Res}
\begin{figure} 
\includegraphics[scale=0.3]{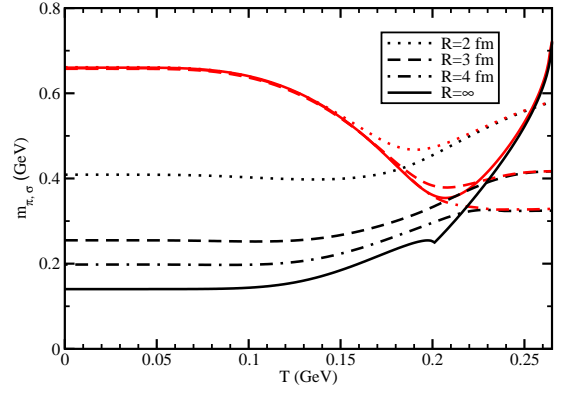}
\caption{(Color online) $T$ dependence of pion (black) and sigma (red) meson masses for 
$R=\infty$ (solid line), $4$ fm (dash-dotted line),
$3$ fm (dash line) and $2$ fm (dotted line).} 
\label{mps_T}
\end{figure}
\begin{figure} 
\includegraphics[scale=0.3]{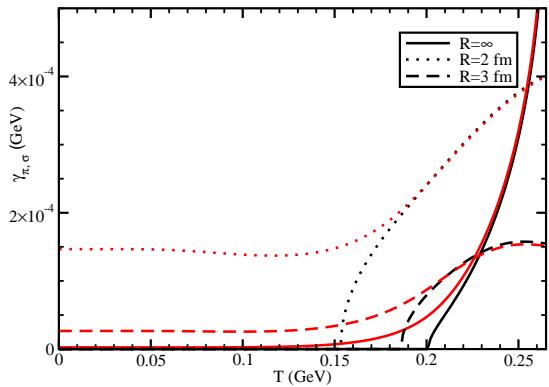}
\caption{(Color online) $T$ dependence of pion (black) and sigma (red) meson decay widths 
through quark-anti-quark channel for different system size.} 
\label{Gps_T}
\end{figure}
\begin{figure} 
\includegraphics[scale=0.3]{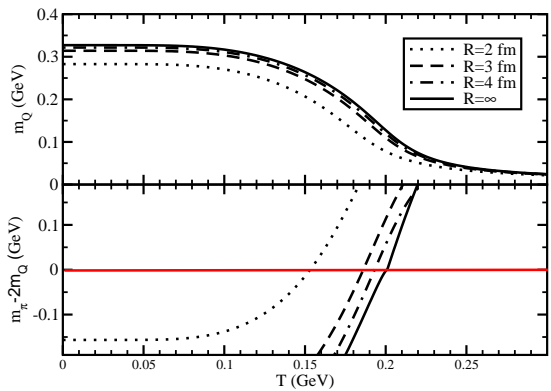}
\caption{$T$ dependence of (a) $M_Q$ and (b) $m_\pi-2M_Q$ for different system size.
Straight horizontal red line, located at $m_\pi-2M_Q=0$, are indicating corresponding Mott 
temperature $T_M$ for different values of $R$.} 
\label{MQ_Mott_T}
\end{figure}
\begin{figure} 
\includegraphics[scale=0.3]{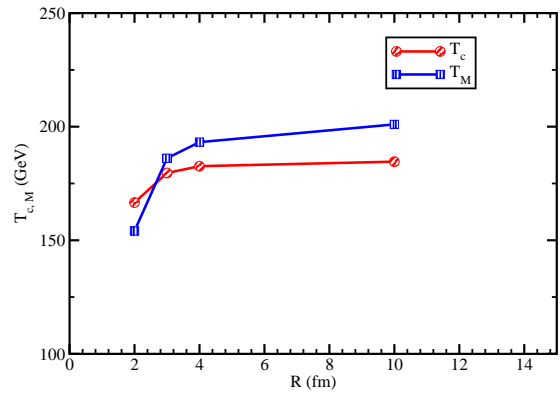}
\caption{Modification of transition temperature $T_c$ and Mott temperature
$T_M$ for changing system size $R$, where $R=10$ is approximately considered as infinite volume.} 
\label{Tc_R}
\end{figure}
In this section we will present the numerical results for the properties of the $\pi$ and $\sigma$ mesons in a hot and dense 
environment. The masses of the $\pi$ and $\sigma$ mesons are obtained from Eq.~(\ref{polemass}) and the decay widths 
can be calculated from Eq.~(\ref{decay-width}). 

Fig.~(\ref{mps_T}) shows the temperature dependence of chiral partners $\pi$ and $\sigma$ mesons for different
system size $R$. As we know that in our real world (for $T=0$) their masses are well separated but at high temperature they
will be in degenerate states. This fact of merging chiral partners is considered as alternative signature of chiral symmetry
restoration. This transition from chiral symmetry breaking to restoration is noticed for both infinite and finite
system size but their merging pattern become different. At $T=0$, mass of $\pi$ meson enhances as we decrease the system size,
while mass of $\sigma$ meson remain unchanged. Actually as the volume decreases, the difference between $\pi$ and $\sigma$
meson also decreases, which indicates that the chiral symmetry effect reduces with decreasing volume. 
Also $T_C$ changes with 
the volume of the system and shifted to lower temperature for smaller system size.

Next, in Fig.~(\ref{Gps_T}), we have plotted $\pi$ and $\sigma$ decay widths in $Q{\bar Q}$ channel for different system sizes.
Since $\sigma$ meson mass always remain greater than two times of quark mass ($m_\sigma>2M_Q$) in entire temperature range,
therefore we will get non-zero $\gamma_\sigma$ in entire $T$. Whereas, for the case of $\pi$ meson mass, the kinematic threshold
$m_\pi>2M_Q$ is valid above the Mott temperature $T_M$, below which the $\pi\rightarrow Q{\bar Q}$ decay is forbidden.
For $R=\infty$ case, $T_M=0.201$ GeV and one can find that black solid line, denoting $\gamma_\pi$, has started to be non-zero
beyond that temperature. Similar to mass merging, the decay width merging of chiral partners are also seen. 
It is expected as
the kinematic phase-space of two decay probabilities depend on their mass only. 
Their coupling constants are not different as
in NJL model, $\pi$ and $\sigma$ meson states are basically
considered as condensates of same quark composition but with different 
spin-parity quantum number $J^\pi$ (=$0^-$ and $0^+$ for $\pi$ and $\sigma$
meson respectively). 
When we consider finite size effect, we find that the $\gamma_\sigma$ is getting enhancement in low $T$ domain.
Mott temperature is decreasing as $R$ decreases, which can be seen
from shifting threshold pion decay width along $T$-axis.

Fig.~\ref{MQ_Mott_T}(a) shows quark masses as a function of $T$ and $R$. Knowing $M_Q(T,R)$ and $m_\pi(T,R)$, one can find
the Mott temperature $T_M(R)$ as a function of $R$, where $m_\pi-2M_Q=0$. It is plotted in Fig.~\ref{MQ_Mott_T}(b), from
where we get $T_M(R)$, which is plotted by red solid line with circles in the Fig.~\ref{Tc_R}. From Fig.~\ref{MQ_Mott_T} one 
can see that the Mott temperature value decreases for lower system size. The value of the Mott temperature for $R=2 fm$ is 
quite smaller than the higher volume systems. 
%

Fig.~\ref{Tc_R} represents the variation of the transition temperature and Mott transition temperature with $R$. The transition 
temperature can be obtained from the maxima of the fist derivative of the chiral condensates for the different finite system
size. As $R$ increases both the transition temperature and the Mott transition temperature increases and after $R=4fm$, both 
the transition temperature and the Mott transition temperature attained saturation. The value of the Mott transition temperature
is lower than the transition temperature below $R=3fm$. As we increase the size of the system the Mott temperature value 
start increasing than the transition temperature and saturates at a higher value than the transition temperature. 

\begin{figure} 
\includegraphics[scale=0.3]{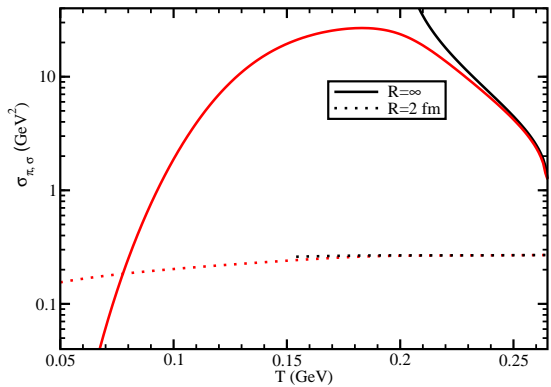}
\caption{(Color online) Conductivity of $\pi$ (black) and $\sigma$ (red) mesons or light flavor condensates with quantum
number $J^\pi=0^-$ and $0^+$ for $R=\infty$ (solid line) and $2$ fm (dotted line).} 
\label{cond_T}
\end{figure}

Using the $m_{\pi,\sigma}(T,R)$ and $\gamma_{\pi,\sigma}(T,R)$ in Eq.~(\ref{cond_ex}), 
we have obtained the conductivity of
the pion and sigma $\sigma_{\pi,\sigma}(T,R)$, which is plotted in Fig.~(\ref{cond_T}).
Below the Mott temperature, the divergence nature of $\sigma_{\pi}(T,R)$ can be clearly seen
due to the relation $\sigma_{\pi}\propto 1/\gamma_{\pi}$.
The conductivity of the pion and sigma
meson changes significantly with the variation of the system size and decreases with the decreasing volume. 
The conductivity
of pion and sigma meson merges after the transition temperature. 
For lower system size the conductivity decreases with the 
decreasing transition temperature and so the pion and sigma meson merges at a lower value. 
\begin{figure} 
\includegraphics[scale=0.3]{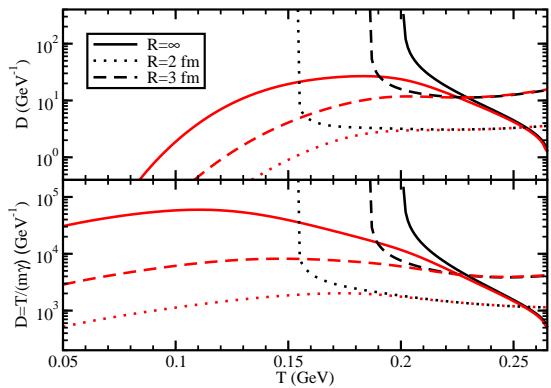}
\caption{(Color online) Diffusion coefficients $D$ of $\pi$ (black) and $\sigma$ (red) mesons or light flavor condensates with quantum
number $J^\pi=0^-$ and $0^+$ for $R=\infty$ (solid line), $3$ fm (dash line) and $2$ fm (dotted line).} 
\label{D_T}
\end{figure}
\begin{table} 
	\begin{center}
	\caption{Transition temperature ($2^{nd}$ column), Mott temperature ($3^{rd}$ column), Temperatures,
	where masses ($4^{th}$ column), drag coefficients ($2^{nd}$ column), diffusion coefficients 
	($3^{rd}$ column) and conductivity ($4^{th}$ column) of chiral partners are merged for 
	$R=\infty$ ($2^{nd}$ and $5^{th}$ rows)
	and $R=2$ fm ($3^{rd}$ and $6^{th}$ rows), temperatures are in MeV.}
	\label{tab:table1}
	\begin{tabular}{ |c|c|c|c|} 
		\hline
		 Size & Transition & Mott & Mass   \\
		 & Temperature & Temperature & Doublet \\ 
		\hline
		$R=\infty$ & 184 & 200 & 245  \\
		\hline
		$R=2$ fm & 166 & 153 & 223  \\
		\hline
		\hline
		 Size & Drag & Diffusion & Conductivity  \\
		 & Doublet & Doublet & Doublet  \\ 
		\hline
		$R=\infty$ & 234 & 245 & 236  \\
		\hline
		$R=2$ fm & 188 & 195 & 193  \\
		\hline
	\end{tabular}
\end{center}
\end{table}
Fig.\ref{D_T}(a) shows the diffusion coefficients of $\pi$ and $\sigma$ mesons by using Eq.~(\ref{D_ex}),
while Fig.\ref{D_T}(b) shows corresponding results by using Einstein relation, given in Eq.~(\ref{Einstein}).
Here also, one can notice the divergence nature of $D_{\pi}(T,R)$ below the Mott temperature
and realize the responsible relation $D_{\pi}\propto 1/\gamma_\pi$.

At the end, if we briefly take a look on all the quantities of chiral partners
for infinite ($R=\infty$) and finite ($R=2$ fm) sizes of medium, then we can 
get a table, given in (\ref{tab:table1}). It clearly implies that transition
temperature, Mott temperature are shifted in lower temperature when we go from 
infinite to finite matter. Table~(\ref{tab:table1}) has also documented the temperatures,
where masses ($4^{th}$ column), drag coefficients ($2^{nd}$ column), diffusion coefficients 
($3^{rd}$ column) and conductivity ($4^{th}$ column) of $\pi$ and $\sigma$ mesons are merged.
Similar to chiral condensate melting, forming mass doublets beyond the transition temperature
is an alternative realization of chiral symmetry restoration. In that regard, merging of other
quantities like drag, diffusion coefficients, conductivity might also be considered as alternative
realization chiral symmetry restoration. In fact, collection all quantities provide a rich 
understanding on the different thermodynamical properties of chiral partners and transition
details from breaking to restore phase of chiral symmetry. Transition point with respect to 
the drag, diffusion coefficients, conductivity will also be shifted in lower temperature, when
one goes from infinite to finite size matter.

\section{Summary} 
\label{sec:sum}
 We have studied the finite volume effect on the spectral functions of strongly interacting matter
 at zero chemical potential. We have shown the pion and sigma meson masses and decay widths at different finite 
 system sizes. Also, we have calculated the conductivity and diffusion coefficients of pion
 and sigma mesons. All the quantities have shown the significant variation with the finite system size.   

 At low temperature the chiral symmetry is broken and after the transition temperature the chiral symmetry is restored.
 The transition from the chiral symmetry broken phase to the chiral symmetry restored phase can be visualized in both
 infinite volume system and the finite volume system. Based on the quark condensate or quark mass melting,
 we can define an chiral transition temperature, which is shifting to lower values when we go from infinite to
 finite size matter. An alternative chiral symmetry restoration can be realized from merging
 of $\pi$ and $\sigma$ masses near and after the transition temperature. This merging point
 is also shifting towards lower temperature due to finite size consideration.
 
 
 We have shown the decay widths of the pion and sigma meson masses with different system size. 
 Decay widths are basically estimated from the imaginary part of self-energy for pion and sigma meson,
 which interpret the thermodynamical probabilities of their dissociation to quark, anti-quark channels.
 For the sigma meson,
 the decay width can be seen for the entire temperature range. However for the pion, the decay width starts after the 
 Mott transition temperature, which also decreases with decreasing system size.
 Similar to masses of the chiral partners, their the decay widths also merge in the temperature domain
 of restored phase.  Again the finite size consideration make their merging points
 shift towards lower values of temperature. 
 For  finite size effect, we find that the decay width of $\sigma$ meson is getting 
 enhancement in low $T$ domain, which is quite interesting and new outcome. 
%
%

  Considering the dissociation process as dragging mechanism of $\pi$, $\sigma$ modes with medium,
  we have estimated their diffusion coefficients and conductivities. Low temperature
  $\pi$ mode diffusion or conduction is diverged due to vanishing drag process below the 
  Mott temperature but its non-divergent values beyond Mott temperature are merged latter
  with corresponding quantities of $\sigma$. Similar to merging of masses and decay widths
  of chiral partners, their merging of diffusion and conduction
  values can be considered as alternative realization of restored phase and their merging
  points again shift towards the low temperature direction when size of the matter is reduced. 

{\bf Acknowledgment:} PD thanks to WoSA scheme of DST funding with grant no....

\end{document}